\documentclass[]{emulateapj}
\newcommand{\dmr}{{\sl DMR }}
\newcommand{\qmap}{{\sl QMAP }}

\newcommand{\msam}{{\sl MSAM }}
\newcommand{\wmap}{{\sl WMAP }}
\newcommand{\wmapc}{{\sl WMAP}}
\newcommand{\maximac}{{\sl MAXIMA}}
\newcommand{\maxima}{{\sl MAXIMA }}
\newcommand{\maximai}{{\sl MAXIMA-I }}
\newcommand{\maximaiic}{{\sl MAXIMA-II}}
\newcommand{\maximaic}{{\sl MAXIMA-I}}
\newcommand{\maximaii}{{\sl MAXIMA-II }}
\newcommand{\boom}{{\sl BOOMERANG }}
\newcommand{\archeops}{{\sl ARCHEOPS }}
\newcommand{\cbi}{{\sl CBI }}
\newcommand{\dasi}{{\sl DASI }}
\newcommand{\vsa}{{\sl VSA }}

\def\bmp{\begin{minipage}}
\def\emp{\end{minipage}}
\def\bfig{\begin{figure}}
\def\efig{\end{figure}}
\def\beq{\begin{eqnarray}}
\def\eeq{\end{eqnarray}}
\def\bi{{\bf (1)}}
\def\bii{{\bf (2)}}
\def\biii{{\bf (C)}}

\def\mathrelfun#1#2{\lower3.6pt\vbox{\baselineskip0pt\lineskip.9pt
    \ialign{$\mathsurround=0pt#1\hfil##\hfil$\crcr#2\crcr\sim\crcr}}}

\newcommand{\abroe}{{M.E. Abroe}}
\newcommand{\borrill}{{J. Borrill}}
\newcommand{\ferreira}{{P.G. Ferreira}}
\newcommand{\hanany}{{S. Hanany}}
\newcommand{\jaffe}{{A. Jaffe}}
\newcommand{\johnson}{{B. Johnson}}
\newcommand{\lee}{{A.T. Lee}}
\newcommand{\rabii}{{B. Rabii}}
\newcommand{\richards}{{P.L. Richards}}
\newcommand{\smoot}{{G. Smoot}}
\newcommand{\stompor}{{R. Stompor}}
\newcommand{\winant}{{C. Winant}}
\newcommand{\wu}{{J.H.P. Wu}}

\newcommand{\spac}{{Space Sciences Laboratory,
University of California, Berkeley, CA, USA}}
\newcommand{\minnesota}{{School of Physics and Astronomy, University of
  Minnesota, Minneapolis, MN 55455, USA}}
\newcommand{\nersc}{{Computational Research Division,
  Lawrence Berkeley National Laboratory, Berkeley, CA 94720, USA}}
\newcommand{\imperial}{{Astrophysics Group, Blackett Laboratory, 
Imperial College, London SW7 2BW, UK}}
\newcommand{\berkeley}{{Department 
of Physics, University of California, Berkeley, CA, USA}}
\newcommand{\pdbl}{{Physics Division,
  Lawrence Berkeley National Laboratory, Berkeley, CA 94720, USA}}
\newcommand{\taiwan}{{Department of Physics, National Taiwan
University, Taipei 106, Taiwan}}
\newcommand{\oxford}{{Astrophysics, University of Oxford, 
Oxford OX1 3RH, UK}}

\begin{document}
\title{Correlations between the \wmap and \maxima Cosmic Microwave
Background anisotropy maps}

\author{
\abroe \altaffilmark{1},
\borrill \altaffilmark{2,3},
\ferreira \altaffilmark{4},
\hanany \altaffilmark{1},
\jaffe \altaffilmark{5},
\johnson \altaffilmark{1},
\lee \altaffilmark{6,7}, 
\rabii \altaffilmark{3,6}, 
\richards \altaffilmark{6}, 
\smoot \altaffilmark{6,7},
\stompor \altaffilmark{2,3},
\winant \altaffilmark{6},
\wu \altaffilmark{8}
}
\altaffiltext{1}{\minnesota}
\altaffiltext{2}{\nersc}
\altaffiltext{3}{\spac}
\altaffiltext{4}{\oxford} 
\altaffiltext{5}{\imperial} 
\altaffiltext{6}{\berkeley}
\altaffiltext{7}{\pdbl}
\altaffiltext{8}{\taiwan}
\email{mabroe@physics.umn.edu}


\begin{abstract}

We cross-correlate the cosmic microwave background temperature anisotropy 
maps from the \wmapc, \maximaic, and \maximaii experiments.  We use
the cross-spectrum, which is the spherical harmonic transform of the 
angular two-point correlation function, to quantify the correlation as a function of 
angular scale. We find that the three possible pairs of cross-spectra
are in close agreement with each other and with the power spectra of
the individual maps. The probability that there is 
no correlation between the maps is smaller than $1\times10^{-8}$. We 
also calculate power spectra for maps made of differences between 
pairs of maps, and show that they are consistent with no signal. 
The results conclusively show that the three experiments not only 
display the same statistical properties of the CMB anisotropy,
but also detect the same features wherever the observed sky areas
overlap. We conclude that the contribution of systematic errors
to these maps is negligible and that \maxima and \wmap have
accurately mapped the cosmic microwave background anisotropy.
   
\end{abstract}

\section{Introduction}

Temperature fluctuations in the cosmic microwave background (CMB)
encode a vast amount of cosmological information about our universe.
CMB photons released from the primordial plasma at the time of recombination
approximately 380,000 years after the Big Bang,
provide thus a picture of the universe in its infancy only somewhat modified by
low-redshift effects such as reionization.
Recently \wmap produced a 13$'$ full sky
measurement of the CMB temperature anisotropy
\citep{bennett/etal:2003}.  This map has been used in conjunction with
other CMB and cosmological data to constrain a number of cosmological
parameters to unprecedented accuracy
\citep{spergel/etal:2003}.
Previous to \wmap a number of experiments produced high quality maps
of CMB temperature anisotropy.  These included both balloon borne bolometric
experiments such as
\boom \citep{debernardis/etal:2000,ruhl/etal:2002},
\maximai \citep{hanany/etal:2000,lee/etal:2001},
and \archeops \citep{benoit/etal:2003}, and ground 
based interferometric experiments, 
\cbi \citep{padin/etal:2001,mason/etal:2003}, 
\dasi \citep{halverson/etal:2002}, and \vsa \citep{grainge/etal:2002}.
Tight constraints were placed on cosmological parameters from
these experiments as well 
(e.g. 
\cite{jaffe/etal:2001,netterfield/etal:2002,
stompor/etal:2001,abroe/etal:2002,pryke/etal:2002}).

\begin{figure*}[th]
\label{fig:wmap_maxima_ps}
\centering
\includegraphics[width=3.5in,angle=90]{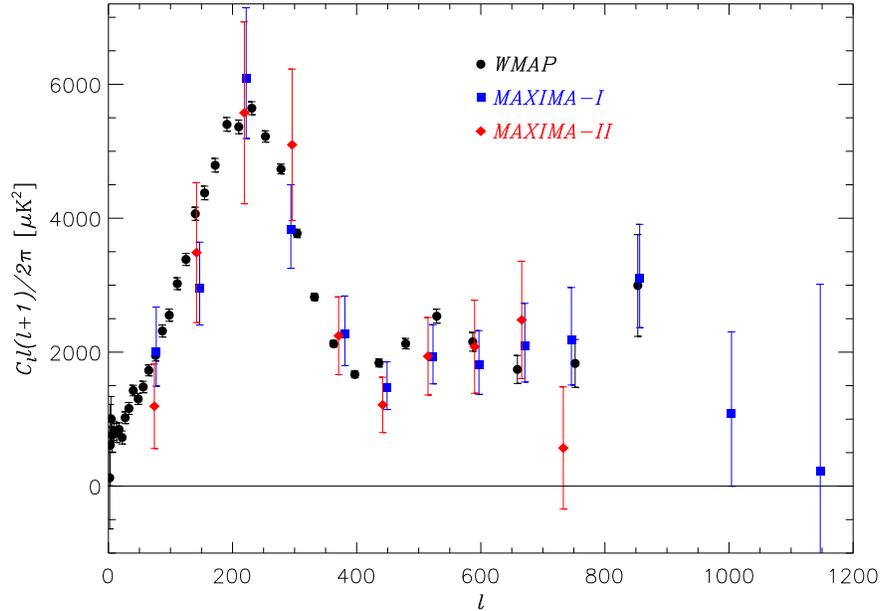}
\caption{
The CMB angular power spectra as measured by
\wmap (\cite{hinshaw/etal:2003}, black circles), 
\maximai (\cite{lee/etal:2001}, blue squares),   
and \maximaii (red diamonds).  
No adjustments
have been made to the calibration for any of the spectra.
For the analysis in this paper we use the version of the 
\maxima-1 data published by \citet{hanany/etal:2000}
(see Section~\ref{sec:maps}).}
\end{figure*}

Given the longer observations and higher sensitivities
of recent CMB experiments the theoretical analysis is more likely
to be limited by systematic rather than statistical errors.
It is therefore important to ensure that systematic errors in these experiments
are sub-dominant compared to statistical errors.
A comparison of the power spectra from the experiments can 
provide some confidence that systematic errors are not dominant.
For instance the power spectra of \wmapc, \maximaic, and \maximaii 
are shown in Figure~\ref{fig:wmap_maxima_ps}. Note that no calibration 
adjustments have been made to the data.
For experiments which observe overlapping parts of the sky
the close agreement of the power spectra 
does not necessarily imply that the spatial 
fluctuations detected by the experiments are identical.
In such cases, and particularly when the experiments have similar 
angular resolution, a more stringent test for systematic errors 
is to cross correlate the temperature fluctuations of one map with the
fluctuations in the other. Positive correlations between temperature anisotropy 
maps would also enhance the confidence in the reconstruction of the spatial
pattern of the CMB. 
Some difficulties arise if the two experiments
under consideration have different pixel resolutions and beam profiles, which
is usually the case.  In that case a straightforward pixel to pixel
comparison is no longer accurate because the CMB signal contained in
corresponding pixels is not the same, and a more elaborate
technique needs to be employed. 

In this paper we use the cross-spectrum as a technique to 
cross-correlate the maps of 
\wmap and \maximac. This is the first reported
cross-correlation of a CMB map with the map of \wmapc, and 
the first release of data from the \maximaii flight.

Other comparisons of CMB anisotropy maps from different
experiments have been performed in the past (e.g.
\citet{ganga/etal:1993}, Tenerife and \dmr \citep{linweaver/etal:1995},
\msam and Saskatoon \citep{knox/etal:1998}, and
\qmap and Saskatoon \citep{xu/etal:2002}).  However, none
of these analyses used the cross spectrum as a technique
for quantifying the amount correlation between data sets.

This paper is organized as follows: 
in Section~\ref{sec:method} we give details
of the cross spectrum and its use for quantifying the amount
of correlations between two CMB anisotropy maps at various angular
scales. 
In Section~\ref{sec:maps} we discuss the maps used in our analysis.
Results of the calculation of the power and cross-spectra for the 
maps, and an analysis of the difference maps
are given in Section~\ref{sec:results}.
We discuss the results
in Section~\ref{sec:discussion}, and summarize in 
Section~\ref{sec:summary}.

\section{Method}
\label{sec:method}

Perhaps the simplest way to compare quantitatively two maps is to
calculate a $\chi^2$ for their difference.  This statistic has the
limiting feature that only overlapping sections of the maps can be
included, and the maps must have identical pixelizations.
Additionally, if the maps have different beam window functions, then
the signal in corresponding pixels is
different and the distribution of the $\chi^2$ statistic would no
longer be $\chi^2$ distributed. The statistical interpretation 
of the $\chi^2$ value would therefore be difficult.

There are a number of statistics which can be calculated between two
CMB maps to quantify their consistency, e.g. the linear and rank
correlation coefficients \citep{press/etal:NRIC:2e}.  Unfortunately,
these correlation coefficients also suffer from the same difficulties
as the $\chi^2$ statistic.

Several authors derived statistics which do take into account partial
overlapping maps, different pixelizations and beam
profiles. \citet{knox/etal:1998} derive both Bayesian and frequentist
techniques for correlating CMB maps, and apply these statistics to
data from the MSAM92, MSAM94, and Saskatoon experiments.  They
advocate the calculation of the ``contamination parameter'', which
gives the probability distribution for the magnitude of a signal that
is not common between the two data sets under consideration.  A low
value for the contamination parameter implies that the data sets are
consistent.  \citet{teg:1999} defines a ``null-buster'' statistic
which gives the number of $\sigma$ between the difference of two maps
and a hypothesis of pure noise.  These statistics can be used both as
an internal consistency check between detectors for the same
experiment \citep{stompor/etal:2003} and to compare
maps from different experiments.

In this paper we use the cross-spectrum to quantify the level of
correlations between the data of \wmap and \maxima.  The cross-spectrum
is the spherical harmonic transform of the real space correlation
function for the two maps.  Whereas  other statistics condense the
information about correlations into a single number, and therefore
result in some loss of information, the cross-spectrum retains more
information by analyzing the correlations as a function of angular
scale.

We now discuss the method for estimating the cross-spectrum from CMB
temperature anisotropy maps.  Consider two maps of the CMB called
$T^\bi$ and $T^\bii$, respectively.  Then \beq \Delta T^\bi_i &=&
s^\bi_i + n^\bi_i \label{eq:t1} \\ \Delta T^\bii_i &=& s^\bii_i +
n^\bii_i, \label{eq:t2} \eeq where $i$ is a pixel index, $s^\bi_i$ and
$s^\bii_i$ are the CMB signal in pixel $i$, and $n_i^\bi$ and
$n_i^\bii$ are the pixel noise for the first and second map,
respectively.  Let the number of pixels in the first and second maps
be $N_{\rm p}^\bi$ and $N_{\rm p}^\bii$, respectively.  We write the
data vector in pixel space as \beq d&=&\left [ \begin{tabular}{c}
$\Delta T^\bi$ \\ $\Delta T^\bii$\\
 \end{tabular}\right],
\eeq where $d$ is now a column vector of length $N_{\rm p}^\bi+N_{\rm
p}^\bii$.  Assuming the signal and noise within each experiment are
uncorrelated and that the noise between experiments is uncorrelated,
and using Equations~\ref{eq:t1} and \ref{eq:t2} we find that \beq
\langle d d^T \rangle &=&M\equiv \left[\begin{tabular}{ccc} $ S^\bi
+{\cal N}^{\bi} $ &$ S^\biii $\\ $ S^{\biii T} $ &$ S^\bii+{\cal
N}^\bii $\\
\end{tabular}\right],
\eeq
where ${\cal N}^\bi=\langle n^{\bi} n^{\bi T} \rangle$ and 
${\cal N}^\bii=
\langle n^{\bii} n^{\bii T} \rangle$
are the pixel noise covariance matrices for the first and second
experiment, respectively.  The quantities $S^\bi$, $S^\bii$, and $S^\biii$ are the
CMB signal covariance matrices, which can be written as
\beq
 S^\bi_{ij}&=&\sum_\ell{{2\ell+1}\over{4\pi}} 
C_\ell^\bi B_\ell^{\bi 2} P_\ell(\cos\theta_{ij}) \\
 S^\bii_{ij}&=&\sum_\ell{{2\ell+1}\over{4\pi}} 
C_\ell^\bii B_\ell^{\bii 2} P_\ell(\cos\theta_{ij}) \\
 S^\biii_{ij}&=&\sum_\ell{{2\ell+1}\over{4\pi}} 
C_\ell^\biii B_\ell^\bi B_\ell^\bii P_\ell(\cos\theta_{ij}),
\eeq
where $B_\ell^\bi$ and $B_\ell^\bii$ are the beam profiles
for the first and second experiment, $P_\ell$ are Legendre polynomials,
and $\theta_{ij}$ is the angle between pixels $i$ and $j$.
The auto-spectra $C_{\ell}^\bi$ and $C_{\ell}^\bii$ are commonly called
the power spectra for the first and second map, respectively, and 
$C_{\ell}^\biii$ is defined as the cross-spectrum. 
In the zero noise case the cross-spectrum is limited by
the requirement that $C_\ell^{\biii} \le \sqrt{C_\ell^\bi C_\ell^\bii}$.
Otherwise $M$ would have negative eigenvalues and therefore
be unphysical.
Up to an irrelevant additive constant, the likelihood for $d$ is 
\beq
-2\ln {\cal{L}}&=& (N_p^\bi+N_p^\bii)\ln |M|+
 d^T M^{-1}d  \label{eq:like_pix}.
\eeq
We maximize the likelihood $\cal L$ as a
function of the two auto-spectra and the cross-spectrum.  If two CMB
maps have a high degree of correlation, then we expect the
cross-spectrum to resemble the auto-spectra of both maps.  If the signal in the two
maps is not correlated the cross-spectrum should be consistent with
zero at all angular scales.

To maximize $\cal L$ we adopt a Newton-Raphson technique for finding
the zero of the first derivative of the log likelihood function.  We
use a technique similar to the one described by \citet{bjk:1998}
except that we use the full curvature matrix instead of the Fisher
matrix when calculating the steps for convergence of the
Newton-Raphson algorithm \citep{hm:2002}.  We found that use of the Fisher matrix gave
non-positive definite pixel-correlation,
presumably because the likelihood function of the three spectra
together has a complicated structure.

The cross-spectrum method involves estimating all three power spectra
simultaneously and gives rise to correlations between the different
spectra. Therefore an auto-spectrum estimated for any one experiment
alone may differ from the auto-spectrum calculated when estimating the
cross spectrum.  For the case of a full sky coverage one can show that
the level of correlations between the spectra is proportional to the
amplitude of the cross-spectrum, that is to the amount of common sky
signal \citep{kam/etal:1996}. We therefore expect a high level of
correlations between the spectra if they share the same sky signal at
low $\ell$'s where the contribution of instrument noise is smaller
compared to sample variance.
\begin{figure}[th]
\centering
\includegraphics[width=2.5in,angle=90]{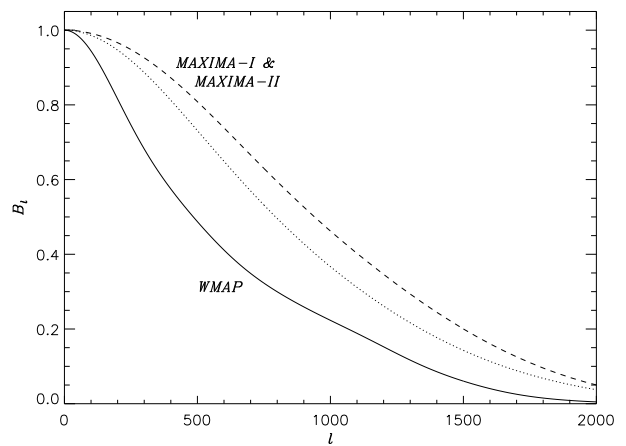}
\caption{The beam filter functions for \maximai (dotted line), 
\maximaii (dashed line), and \wmap (solid line)
considered in this analysis.  
The horizontal
axis is spherical harmonic multipole as a function of $\ell$. 
}
\label{fig:beam}
\end{figure}

The cross-spectrum is a powerful technique for comparing two CMB
anisotropy maps because it accounts for different beam shapes, pixel 
resolutions and sky coverage in a simple and straightforward way.  
For example, one can compute the cross-spectrum for two maps that do not
overlap at all. In such a case the results would not be sensitive to
correlations on small angular scales.  Also, though our formalism
describes estimating the cross-spectrum for two maps,
a further generalization
could be made to estimating correlations between three or
more maps simultaneously.
However, given numerical subtleties which are subsequently 
discussed in Section~\ref{sec:comp_issues}, we found
it prudent, while equally convincing,
to estimate the correlations for only pairs of maps.

The power spectra estimated from the \wmap maps are
actually the cross-spectra from different detectors from the same
frequency band \citep{hinshaw/etal:2003}.  
Our approach is distinct from the one used by the \wmap team
in that they use 
a frequentist approach \citep{hivon/etal:2002} to estimate the
cross-spectra between various detectors of the same frequency band
independently from the auto-spectra.  The approach we use in this 
paper is entirely Bayesian.

\begin{figure*}[t]
\centering
\includegraphics[width=3.5in,angle=0]{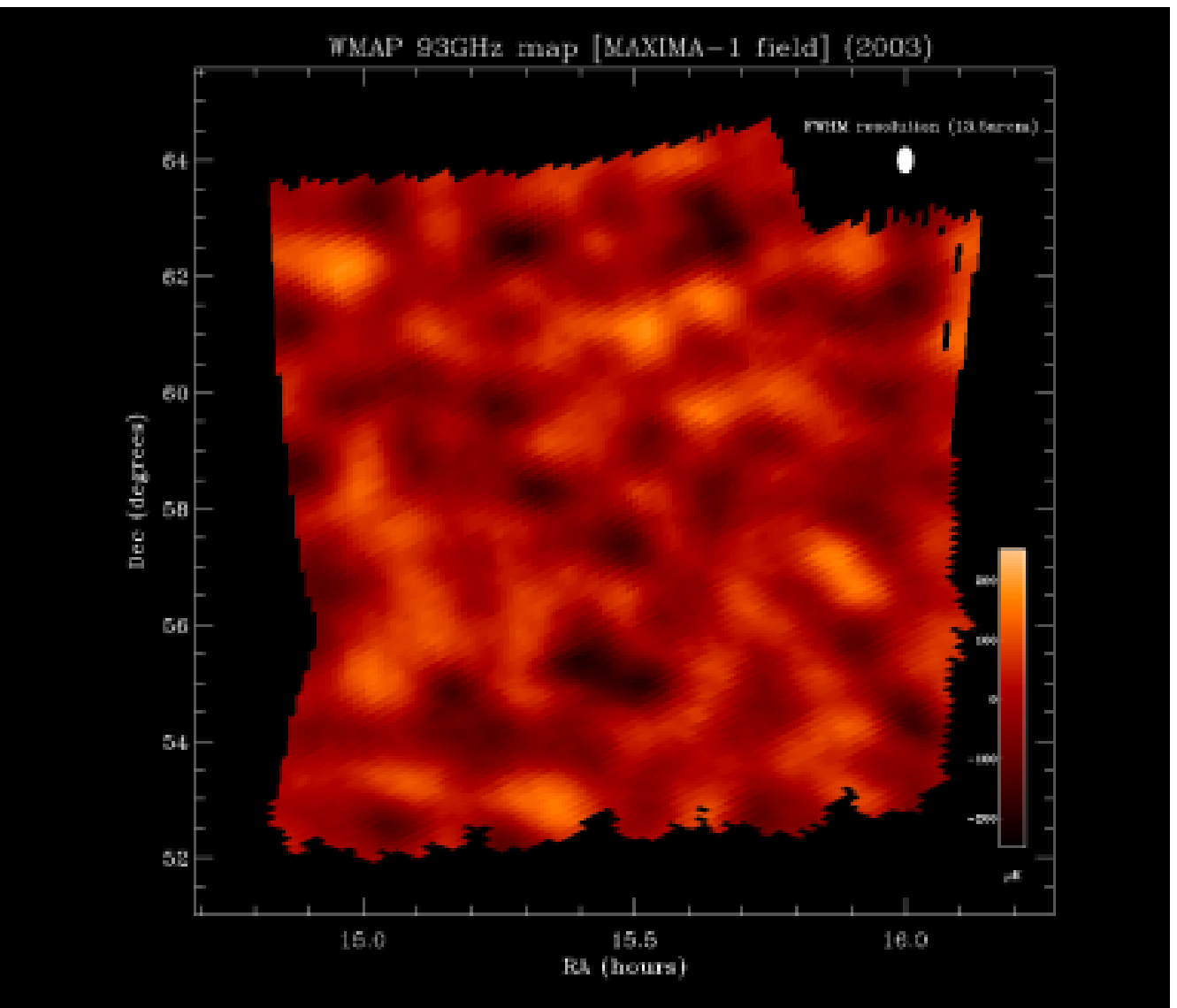}
\includegraphics[width=3.5in,angle=0]{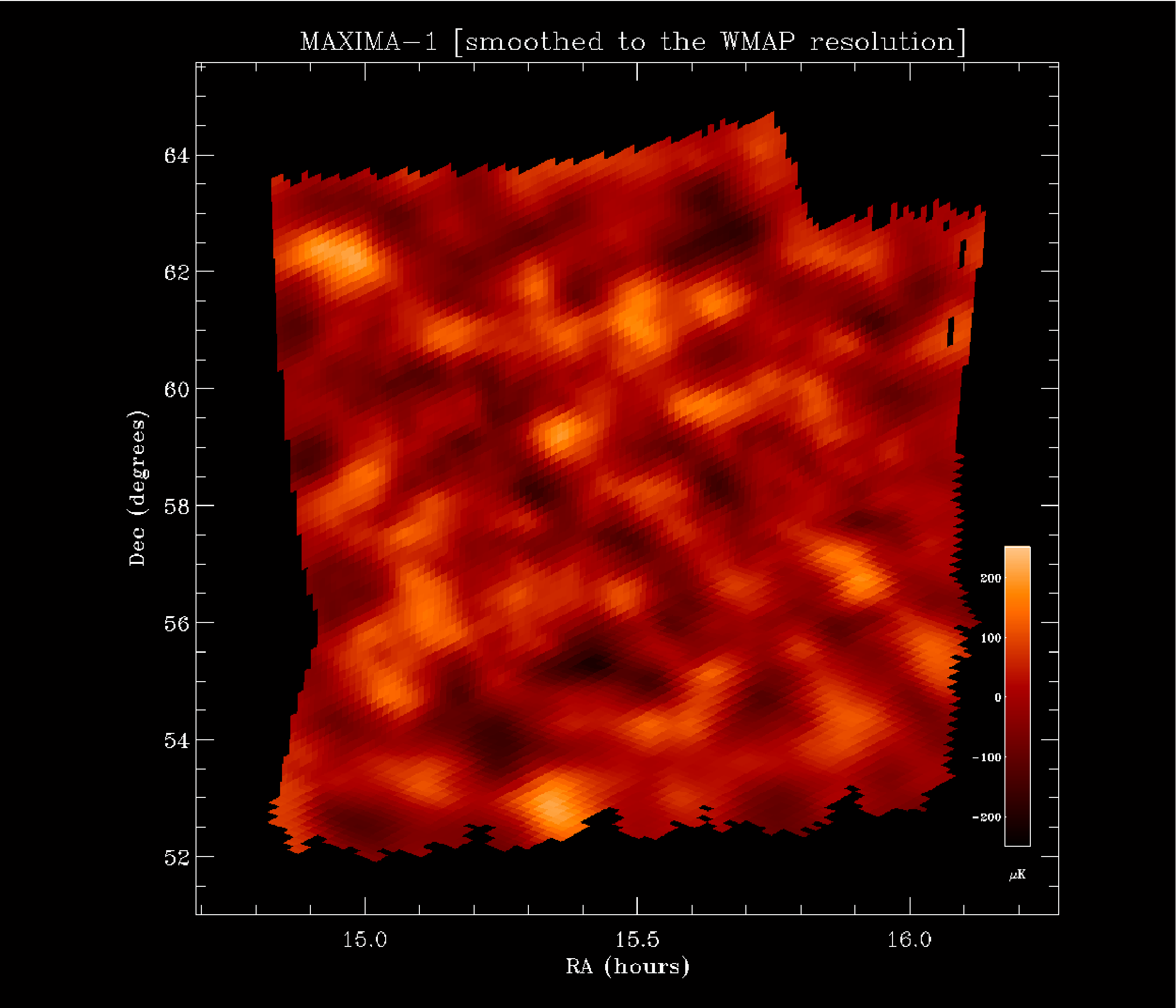}
\includegraphics[width=3.5in,angle=0]{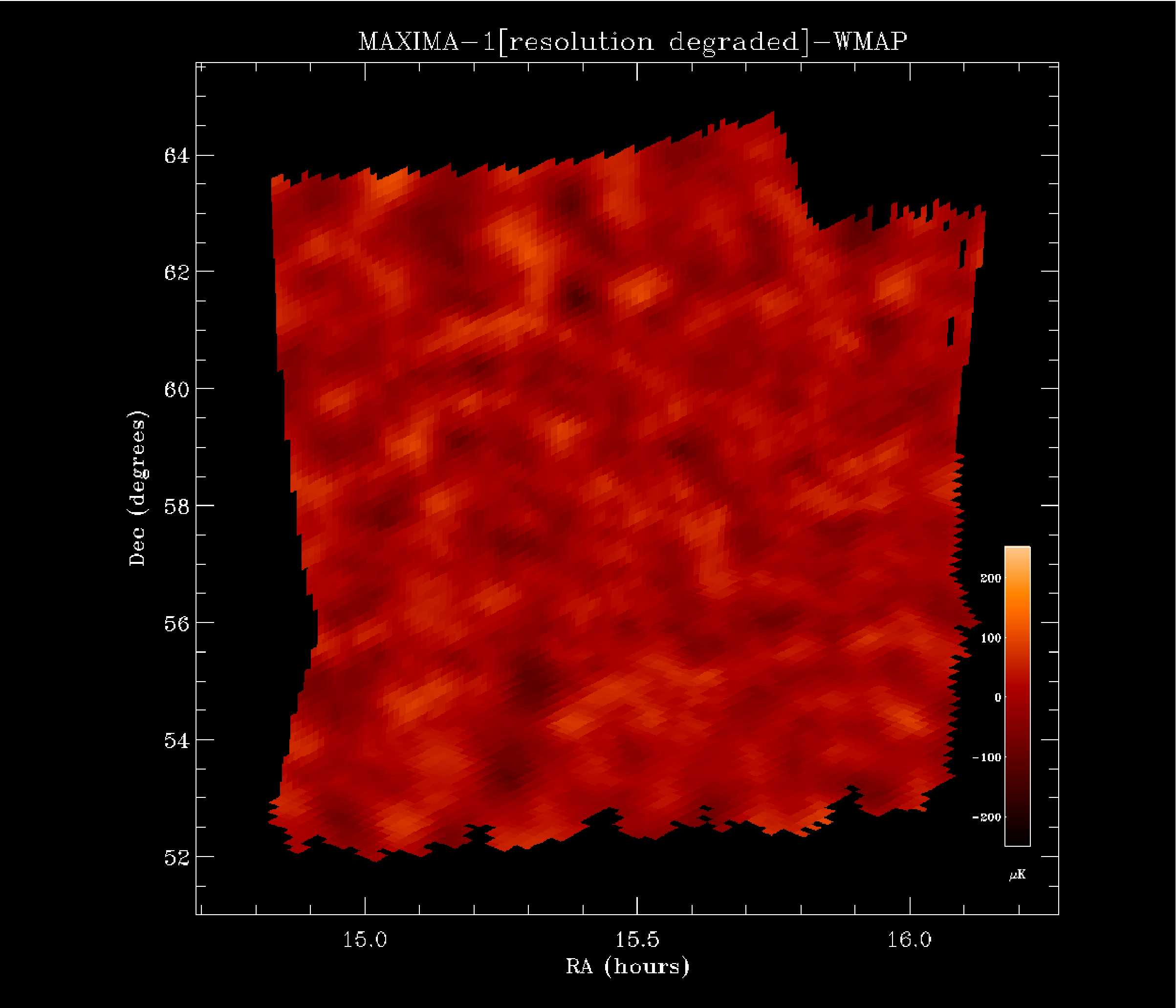}
\caption{
  A comparison of the overlap region of the 
\maximai (upper right panel) and the \wmap 93 GHz (upper left panel) CMB temperature
anisotropy maps, and their difference map (lower panel).  The maps
have been Wiener filtered for a visual comparison; the non-filtered
versions are used in the analysis.
All modes
with $\ell\le 35$ have been removed from these maps.  
The \maximai map, which originally had a resolution of 10$'$, has
been smoothed to the \wmap resolution.
The color scale is from $-250$ to $250$ $\mu{\rm K}$.}
\label{fig:max1_wmap}
\end{figure*}

\begin{figure*}
\centering
\includegraphics[width=3.3in,angle=90]{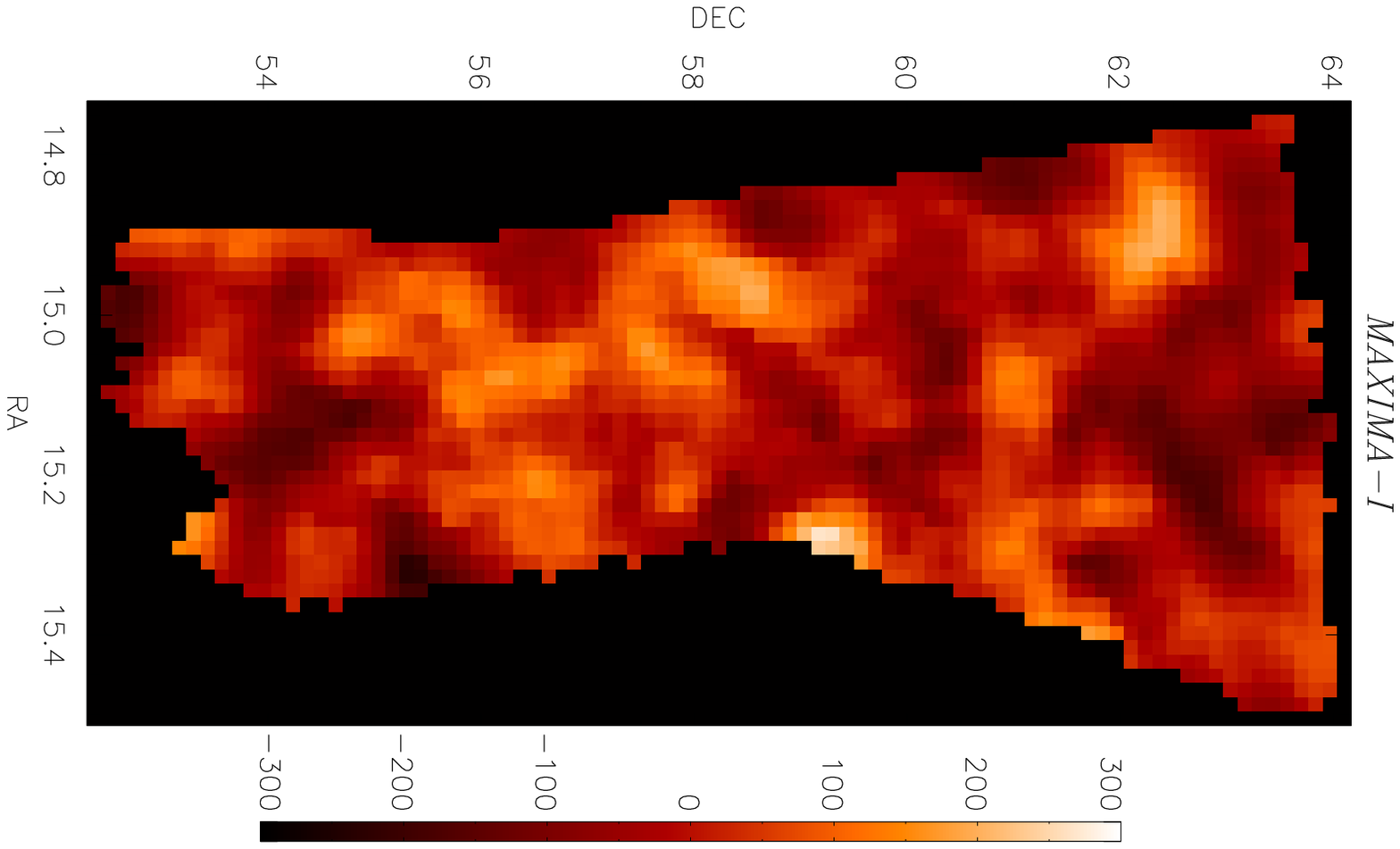}
\includegraphics[width=3.3in,angle=90]{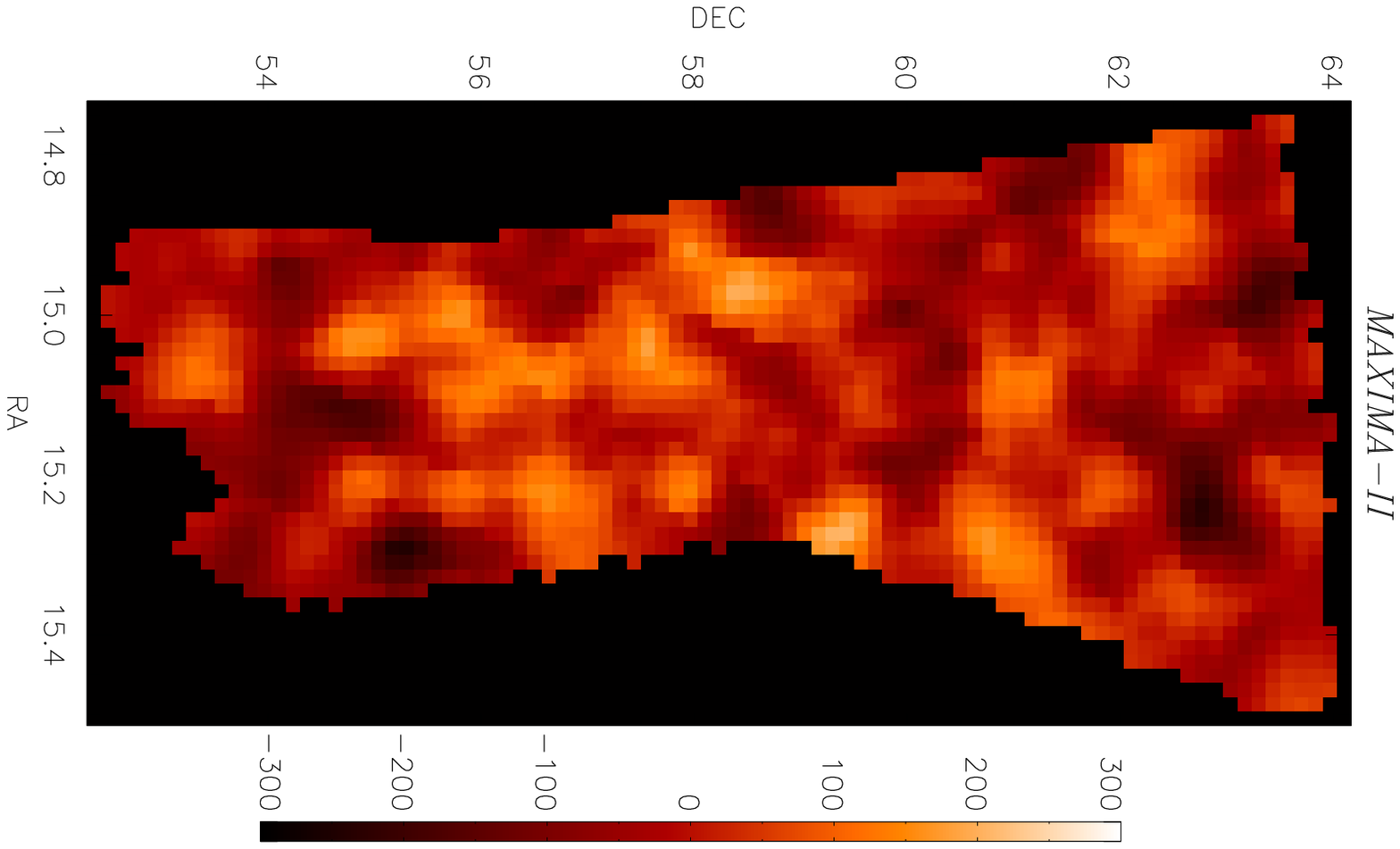}
\includegraphics[width=3.3in,angle=90]{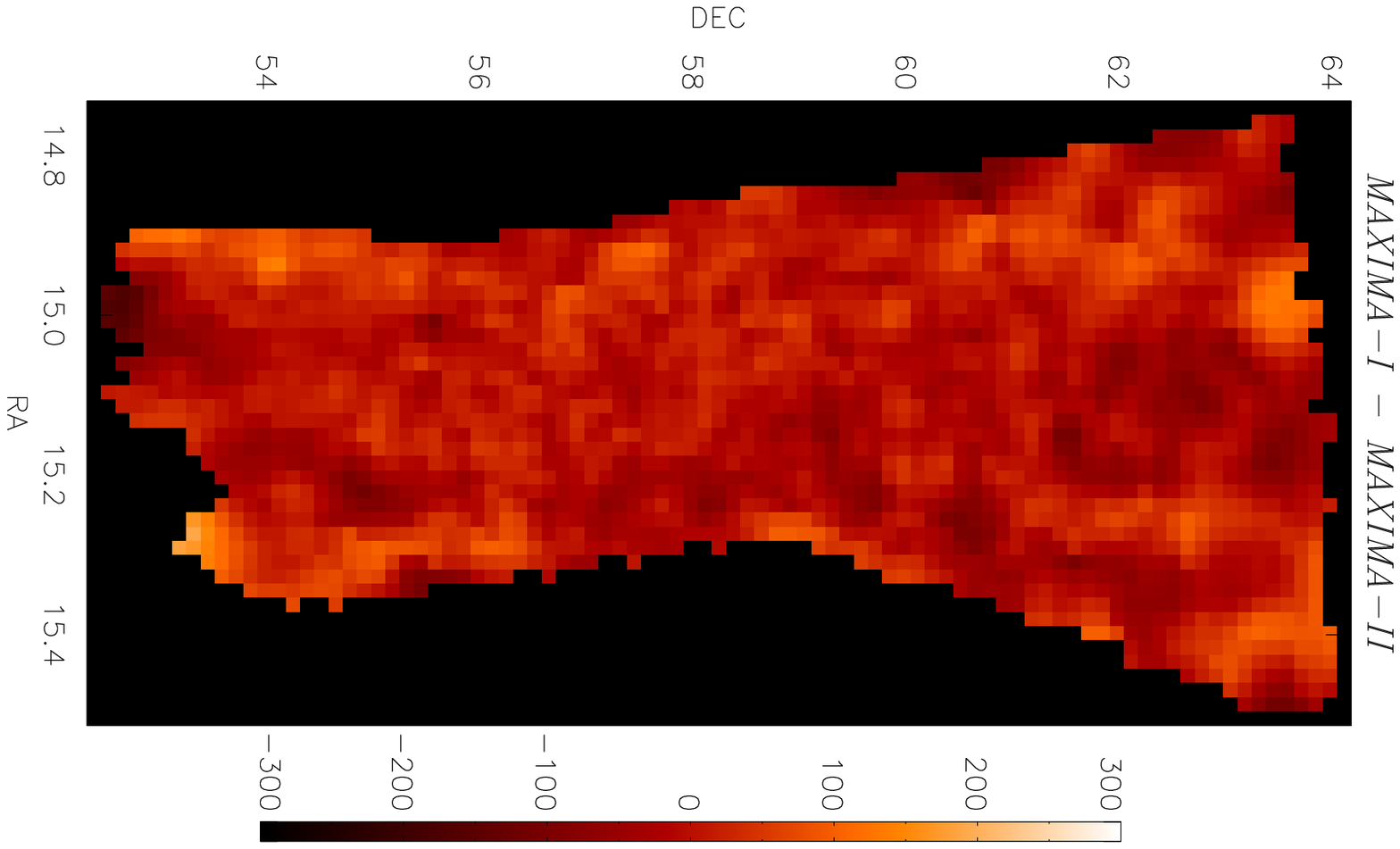}
\caption{The overlap sections of the \maximai and \maximaii
Wiener filtered maps, shown in the left and center
panel, respectively.  The difference
of the Wiener filtered maps is shown in the right panel.  
The color scales range from $-300\mu$K to $300\mu$K.  
Note that while only the overlap section
of the \maximai map is shown here, we use the entire \maximai map for
all analyses.}
\label{fig:maps}
\end{figure*}

\section{The Maps}
\label{sec:maps}

We cross-correlate CMB temperature anisotropy maps from the \wmapc,
\maximai and \maximaii experiments.  The \wmap map used is the W-band
(93 GHz) foreground-cleaned map, and only the portion that overlaps
with the \maximai field.  By analyzing\\ \wmap's derived foreground
maps we find that this portion of the W-band map is free of point
sources, and contains a negligible amount of dust, free-free, and
synchrotron
emission\footnote{http://lambda.gsfc.nasa.gov/product/map/m\_products.cfm}.
The \wmap data at 93 GHz has similar angular resolution and is closest
in frequency to the \maxima data.  This map is pixelized using the
HEALPix\footnote{http://www.eso.org/science/healpix/}
\citep{healpix:1999} pixelization in Celestial coordinates with
nside$=512$, and contains 7,926 pixels.  An nside=512 corresponds
roughly to a pixel with a width of 7$'$.  Following
\citet{bennett/etal:2003} we assume that there are no noise
correlations between pixels and that the beam pattern is non-Gaussian,
azimuthally symmetric and has FWHM of 13.2$'$.  We computed the beam
window function by taking the weighted average of the individual beams
for each of the four detectors used to form the final W-band map.  The
resulting beam profile in $\ell$ space is shown in
Figure~\ref{fig:beam}.  The Wiener filtered version of this map is
shown in the left panel in Figure~\ref{fig:max1_wmap}; we use the raw
data in the analysis.  The Wiener filtering is performed with the
corresponding best fit models for each map.  A more detailed
discussion of the \wmap maps is given in \citet{bennett/etal:2003} and
of algorithms used in their computations in
\citet{hinshaw/etal:2003b}.

The \maximac\/ map-making procedure is described exhaustively in
\citet{stompor/etal:2002}.  The \maximai map we use in this analysis
is the 8 arcminute version of the data published by
\citet{hanany/etal:2000}, which covers a larger area of the sky and
has a coarser resolution compared to the data published by
\citet{lee/etal:2001}.  There are a total of $5,972$ pixels, and this
map covers $\sim 100$ ${\rm deg}^2$ on the sky.  It is a combination
of three $150$ GHz photometers and one 240 GHz photometer.  The beams
for each photometer have a FWHM of $\sim10'$ \citep{hanany/etal:2000}
and the effective window function is calculated using the technique
described in \cite{proty/etal:2001}.  The \maximai map shown in
Figure~\ref{fig:max1_wmap} is Wiener filtered and smoothed to a \wmap
resolution.  The raw version of the map is used for all quantitative
analyses.  Figure~\ref{fig:max1_wmap} also shows the difference of the
\maximai and \wmap Wiener filtered maps.  The pattern of temperature
fluctuations, which is similar in both maps, disappears in the
difference map.

The \maximaii map comes from a flight of the \maxima payload that took
place on 1999 June 17. We use the data from four photometers at 150
GHz, and only the portion of the \maximaii map that overlaps the map
of \maximaic.  More details about the \maximaii flight, data and maps
are given in Rabii et al. (2003, in preparation), and
\citet{stompor/etal:2003}.  The \maximaii map is pixelized using an
8$'$ square pixelization in celestial coordinates, contains 2,757
pixels, and covers $\sim$ 50 ${\rm deg}^2$ on the sky.  The beam
profile for this map is $\sim10'$ FWHM, and again computed using the
techniques described in \cite{proty/etal:2001}.  The \maximaii power
spectrum shown in Figure~\ref{fig:wmap_maxima_ps} has 10 bins of
$\Delta\ell=75$, extending over the $\ell$ range $35 \le \ell \le
785$.  Figure~\ref{fig:maps} shows the overlap region of the \maximai
and \maximaii maps and the difference map.  Identical temperature
fluctuations that are apparent in each of the maps disappear in the
difference map.

\section{Results}
\label{sec:results}

The auto- and cross-spectra for all combinations of the \wmap and
\maxima maps are shown in Figure~\ref{fig:ps}.  The error bars on the
spectra are the square root of the curvature of the likelihood
function about the maximum likelihood parameter value.  In all cases
we compute the spectra in bins of width $\Delta \ell = 75$, over the
interval $111 \le \ell \le 710$, and marginalize over all modes $ \ell
\le 110$ and $\ell \ge 711$.  The appropriate pixel window functions
for each map were convolved with the beam functions in the analysis.
We found that the cross-spectrum estimator did not converge when the
initial bin was split in two, and this is further discussed in
Section~\ref{sec:discussion}.  In all cases the cross-spectra are
consistent with the auto-spectra giving strong evidence for a
correlation between the maps.

\begin{figure}[th]
\centering
\includegraphics[width=2.5in,angle=90]{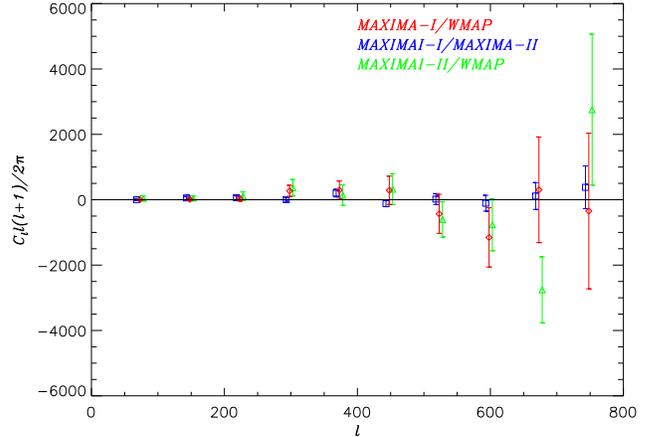}
\caption{
The power spectrum of the
difference maps from all three data sets.  The diamonds (red) are
the power spectrum of the \maximai/\wmap difference map, the squares
(blue) are the power spectrum of the \maximai/\maximaii difference
map, and the triangles (green) are the power spectrum of the
\maximaii/\wmap difference map.  All three spectra are
statistically consistent with the null spectrum.  }
\label{fig:diff}
\end{figure}

We also compute the power spectrum of the difference maps for all
three pairs of maps using bins of $\Delta\ell=75$ over the range
$35\le \ell \le785$.  The \wmap window function is used when computing
the \maximai-\wmap and \maximaii-\wmap difference spectra.  The
expected residual power resulting from different beam profiles is
maximum at the bin centered at $\ell\simeq 300$, and is approximately
equal to the 1$\sigma$ error bar of the \maximai/\wmap difference
power spectrum.  The effect is less than 1$\sigma$ for all remaining
bins.  We use the \maximai window function when computing the
\maximai-\maximaii difference spectrum.  The results are shown in
Figure~\ref{fig:diff}.  Of the 30 band power estimates for the
difference maps, 28 are within 1$\sigma$ of zero power.
\begin{table}[t]
\caption{\small  Cross Spectrum $\chi^2$ Values}
\label{table:chi2}
\small{
\vbox{
\tabskip 1em plus 2em minus .5em
\halign to \hsize {  \hfil#\hfil & \hfil#\hfil & \hfil#\hfil
                  &\hfil#\hfil   \cr
\noalign{\smallskip\hrule\smallskip\hrule\smallskip}
Maps &DoF& $\chi^2_{\rm cutoff}$  & $\chi^2$ \cr
\noalign{\smallskip\hrule\smallskip}
\cr 
\maximai/\wmap     & 8 & 53 & 191   \cr 
\maximai/\maximaii & 8 & 53 & 241    \cr 
\maximaii/\wmap    & 8 & 53 & 150    \cr
\noalign{\smallskip\hrule}
}}}
\small{The $\chi^2$ from Equation~\ref{eq:chi2} calculated for all three
combinations of maps.  
A $\chi^2$ greater than 53
implies that the probability 
that the no correlation hypothesis is true
 is less than $1\times10^{-8}$.}
\end{table}

\begin{table}[t]
\caption{\small  Difference Spectrum  $\chi^2$ Values}
\small{
\vbox{
\tabskip 1em plus 2em minus .5em
\halign to \hsize {  \hfil#\hfil & \hfil#\hfil & \hfil#\hfil
                  &\hfil#\hfil   \cr
\noalign{\smallskip\hrule\smallskip\hrule\smallskip}
Maps &DoF& $\chi^2$   \cr
\noalign{\smallskip\hrule\smallskip}
\cr 
\maximai/\wmap     & 10 & 7.5  \cr 
\maximai/\maximaii & 10 & 8.3   \cr 
\maximaii/\wmap    & 10 & 17.2    \cr
\noalign{\smallskip\hrule}
}}}
\small{The $\chi^2$ of the power spectrum for the difference maps
with the null spectrum.}
\label{table:diff}
\end{table}

To further quantify the level of correlation between the maps we use 
a $\chi^{2}$ statistic to reject the hypothesis that the maps are uncorrelated. 
We write our statistic as
\beq
\chi^2= \sum_{B B'}C_{B}^\biii F_{B B'}C_{B'}^\biii
\label{eq:chi2}
\eeq
where the sum is over band power estimates, and $F$ is the Fisher
matrix for the cross-spectrum.  Because the auto-spectra
and cross-spectrum are estimated simultaneously, we marginalize over the auto
spectra when calculating the $\chi^2$.

To test the null hypothesis we choose a statistical significance
$\alpha=1\times10^{-8}$.  
If $\chi^2$ is greater than the critical
value $53$, 
then the probability that the null hypothesis is
true is less than $1\times 10^{-8}$.
The results are
summarized in Table~\ref{table:chi2}.  In all cases $\chi^2$ is
significantly larger than the critical value giving an essential 
certainty that the no-correlation hypothesis is false.
Note that assuming the $\chi^2$ of Equation~\ref{eq:chi2}
is $\chi^2$ distributed is equivalent to assuming
that the $C_{B}^\biii$ are Gaussian distributed, which
is an approximation.

We also compute the $\chi^2$ of the difference spectra shown in
Figure~\ref{fig:diff} with the null spectra to determine how
consistent these are with no fluctuations in the difference
maps.  The results are shown in Table~\ref{table:diff}.  The 10 power
spectrum bins computed from the \maximai/\wmap and \maximai/\maximaii
difference maps have a $\chi^2$ of 7.5 and 8.3, respectively, with the
null spectrum.  The \maximaii/\wmap
difference map gives a $\chi^2$ of 17.2.  There is a 7\% chance of getting 
$\chi^2\ge17.2$ for 10 DoF.
Overall there is a good fit to the null spectrum model, which
implies that differencing the overlap section of the maps removes the
sky signal and is consistent with noise.

\section{Discussion}
\label{sec:discussion}

\subsection{Auto- and Cross-Spectra}

The auto- and cross-spectra of the different data sets agree with each
other to within $1\sigma$ over almost all $\ell$ bins giving evidence
that at each angular scale all experiments are detecting the same
spatial fluctuations on the sky. All auto- and cross-spectra show the
first acoustic peak in the power spectrum and then a level of power
that is consistent with subsequent peaks.  These results are
consistent with standard inflationary $\Lambda$CDM models.

\begin{figure}
\centering
\includegraphics[width=2.5in,angle=90]{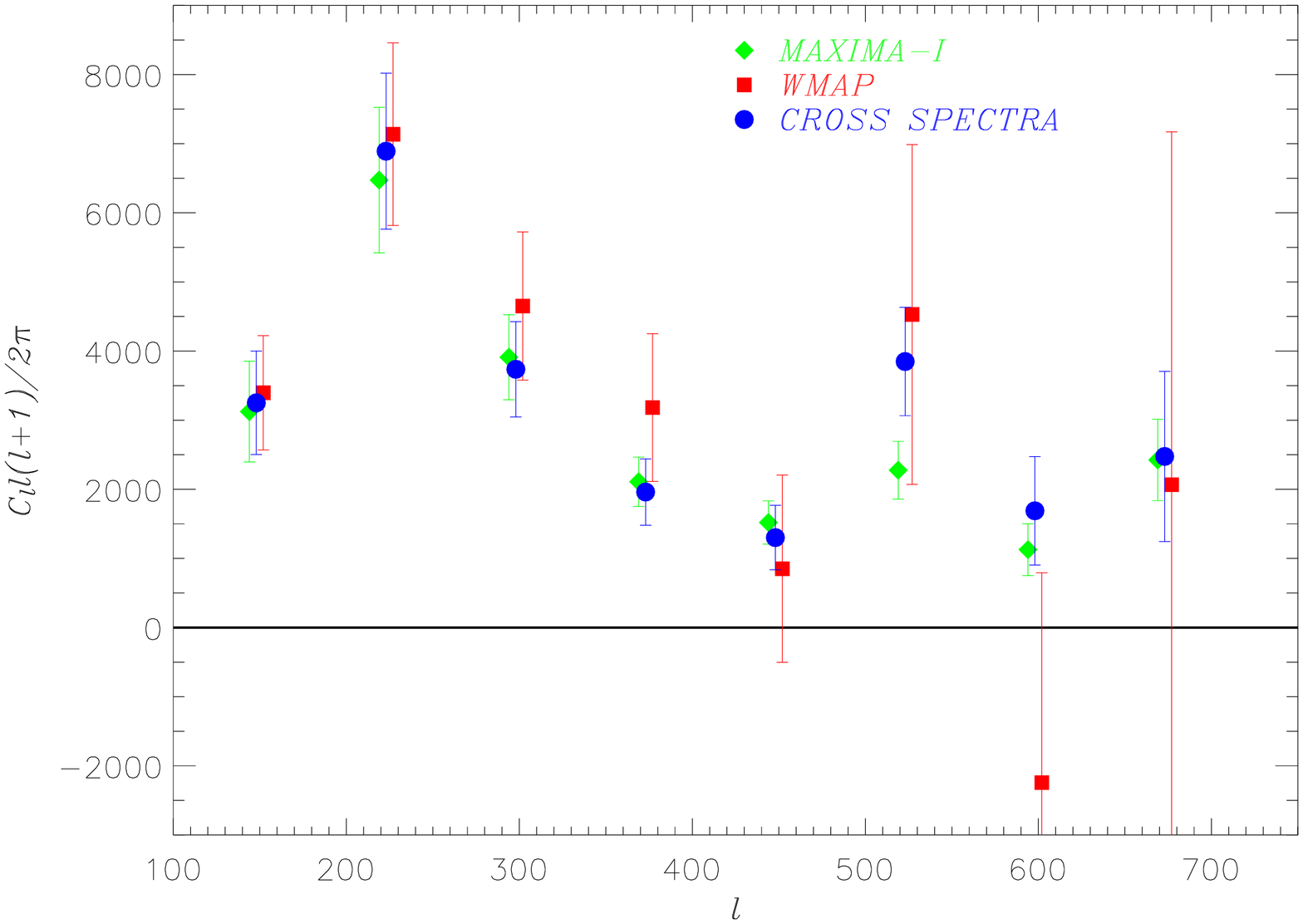}
\includegraphics[width=2.5in,angle=90]{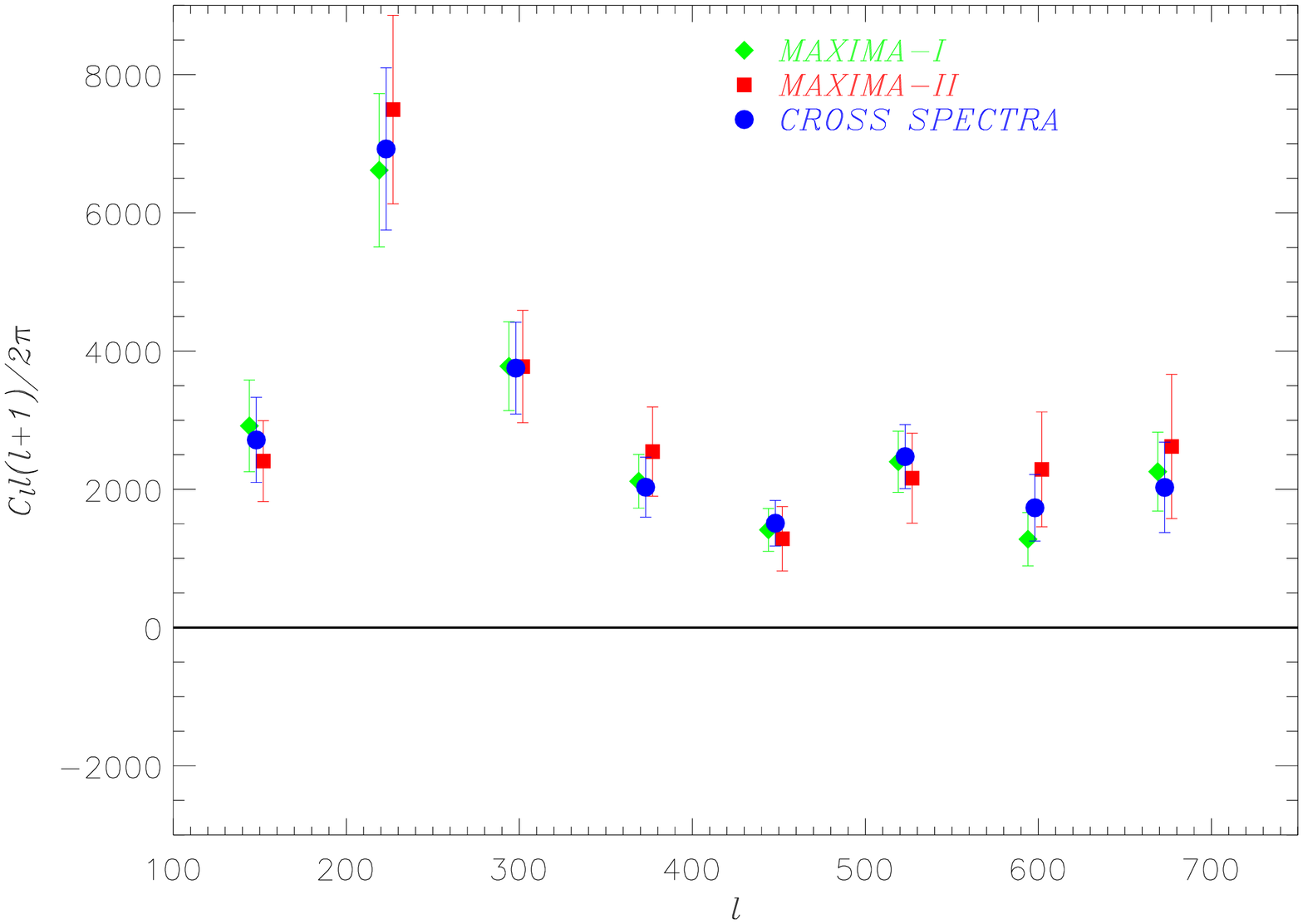}
\includegraphics[width=2.5in,angle=90]{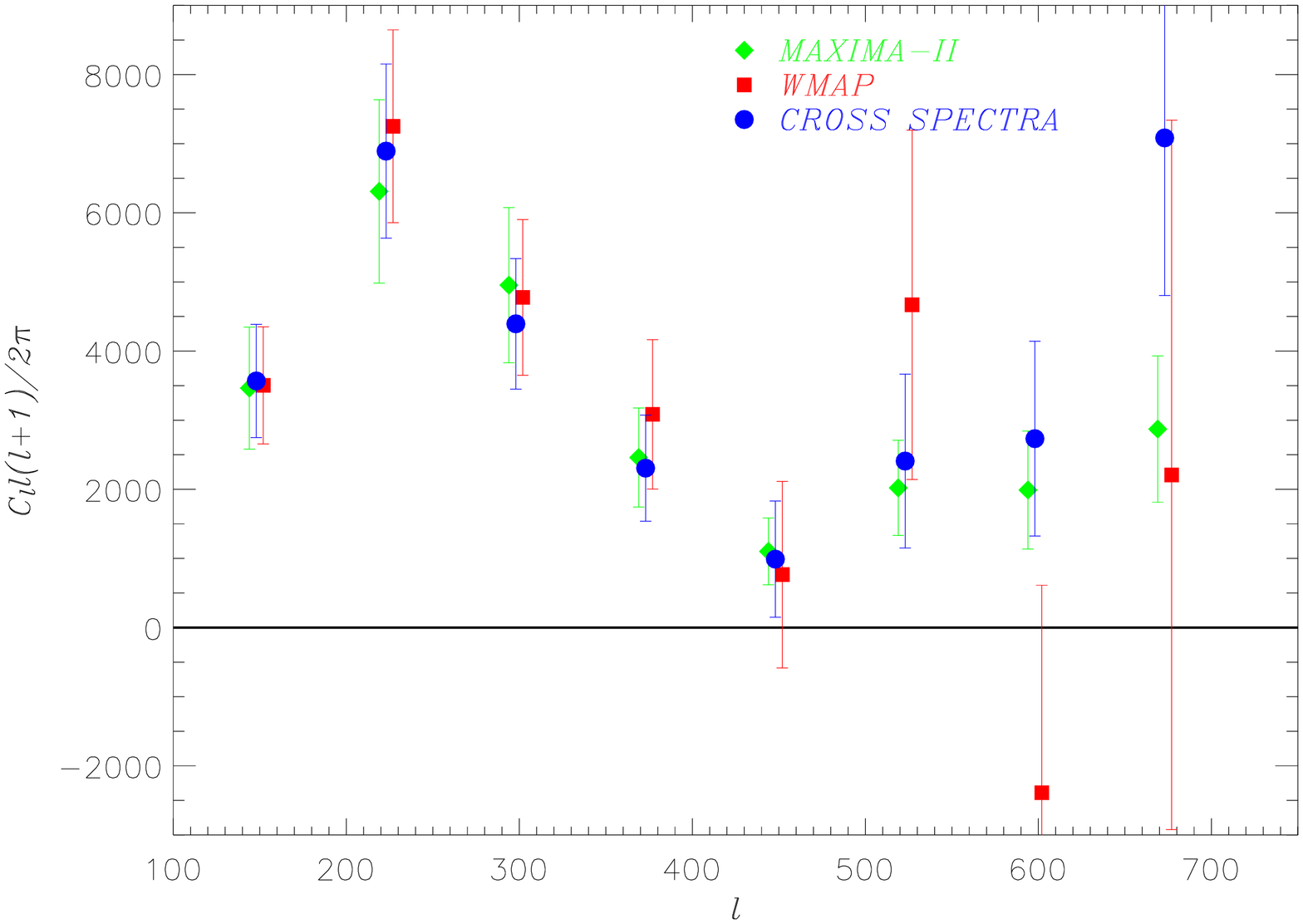}
\caption{The auto-spectra and cross-spectrum estimated for the 
\maxima and \wmap maps.  
From top to bottom:  the results
for \maximai/\wmapc, \maximai/\maximaiic, and \maximaii/\wmap.
Note that only the portion of the \wmap map which overlaps
with the \maximai field is used in the analysis.}
\label{fig:ps}
\end{figure}

Auto-spectra of the overlap section of the \wmap data give increased
error bars at $\ell \ge 486$ because of the limited sky coverage of
the overlap regions, and because of the beam profile of the W-band
map.  We find that the beam pattern alone causes the \wmap
auto-spectrum error bars in the bins
$\ell$=$\{\{486,560\},\{561,635\},\{636,710\}\}$ to be 2-3 larger than
those for \maximai or \maximaiic.  Negative power was found in the bin
$\ell$=$\{486,560\}$ for the \wmap auto-spectra (see the top and
bottom panel of ~\ref{fig:ps}).  There is no requirement that the
auto-spectrum be positive in our estimation method.

A comparison of the auto-spectra shown in Figure~\ref{fig:ps} reveals
that there is a difference between band power estimates for the same
dataset.  This difference arises because the computation of
cross-spectra involves estimating both auto- and cross-spectra
simultaneously, giving rise to correlations between the different
spectra.  The fractional changes in power averaged over bins are 3\%,
6\%, and 15\% for the \wmapc, \maximaic, and \maximaii data sets,
respectively.  If the likelihood distribution of the band powers were
strictly Gaussian, then the maximum likelihood estimates would be the
same regardless of the correlations between spectra.  However, the
likelihood as a function of auto- and cross-spectra is somewhat
non-Gaussian (see Equation~\ref{eq:like_pix}, \cite{bjk:1998}), so the
correlations do effect the band power estimates.  The fact that the
changes between estimates are small suggests, however, that the
distributions are close to Gaussian.

We note that the bin-powers are correlated at the 77\% level or higher
at the lowest $\ell$ bin.  The correlation decreases to less than 10\%
at the highest $\ell$ bin for correlations between the auto-spectra
and to 20\% - 50\% for correlations between the auto- and
cross-spectra. These results are in broad agreement with expectations
given the high amplitude of the cross-spectrum.

\subsection{Computational Issues}
\label{sec:comp_issues}

The strong correlation between the different data sets
leads to some computational difficulties when
attempting to find the maximum likelihood auto- and cross-power
spectra.  As discussed in Section~\ref{sec:method} the
cross-spectrum is limited by the requirement that
$C_\ell^{\biii} \le \sqrt{C_\ell^\bi C_\ell^\bii}$.  We find that using a
quadratic estimator (Fisher matrix) method for calculating the
Newton-Raphson step leads to a mis-estimate of the step for
$\delta C_\ell^\biii$ when starting with a guess significantly far
away from the peak in likelihood space.  This consequently results in
a step which leads to a non-positive definite pixel covariance matrix.
For example, an initial guess of a null spectrum leads to an
unphysical pixel covariance matrix in all three cases we are
considering.  This is remedied by using the curvature matrix to
compute $\delta C_\ell^\biii$, as discussed in
Section~\ref{sec:method}.  Once the parameter values become
sufficiently close to the maximum likelihood values either the
curvature matrix or fisher matrix can be used to find the maximum
likelihood parameters.  Both techniques converge to the same set of
parameters for all three analyses.

The power spectra shown in Figure~\ref{fig:ps} have been calculated
with a broad initial bin at $\ell=\{2,110\}$.  This is because the Newton-Raphson 
likelihood maximization technique did not converge if this low $\ell$ bin 
was split to two.  Some binning
structures would cause negative eigenvalues in the curvature matrix of
the parameters or steps in parameter space that would lead to a
non-positive definite pixel covariance matrix, both of which are
unphysical.

We carried out simulations and found a similar phenomenon.  The
cross-spectra of uncorrelated maps or maps with a small value for the
ratio of expected cross-spectrum to auto-spectrum converged to the
expected answer.  The calculation of the cross-spectrum also converged
with simulated maps that had perfect correlation (i.e. the same map
with different noise realizations) and a broad first bin with 
$\ell=\{2,110\}$. However it did not converge with simulated maps that had perfect
correlation and two bins between $\ell$ of $2$ and $110$.
Therefore, we attribute the computation problems encountered as a
limitation in the method used for computing the cross-spectrum and
not a feature in any of the data sets considered in this analysis.

\subsection{Foregrounds and Systematic Errors}

The cross spectrum and difference spectrum analyses conclusively
demonstrate that all three experiments have mapped the same
temperature fluctuations on the sky.  However, these analyses are not
sensitive to whether the shared fluctuations are CMB in origin or the
result of foreground contamination or a shared systematic effect.

A careful foreground analysis was carried out by both the \wmap and
\maxima teams.  It was shown that the \maximai region of
the sky at 150-240 GHz contains a negligible amount galactic
contamination and that it has no detectable point sources
\citep{hanany/etal:2000,jaffe/etal:2003}.

A detailed analysis of the foreground sources in the \wmap data is
presented in \cite{bennett/etal:2003b}. Although we use \wmap's
foreground-cleaned map,\\ even \wmap's foreground maps have
negligible amount of contamination in the \maximai region.  The RMS
fluctuations in the \wmap 93 GHz dust, synchrotron, and free-free maps
are lower than the corresponding \wmap CMB map by a factor of $17$,
$42$, and $560$, respectively.  Also, the \wmap team find no point
sources in the \maximai region of the sky.

It is unlikely that \wmap and \maxima share systematic errors. 
We therefore conclude that the common signal in the \wmap and \maxima
data is the cosmic microwave background radiation. Since 
the cross-spectra agree with the auto-spectra we conclude that within
the signal-to-noise ratio of the tests systematic errors in the 
data are smaller compared to statistical errors. 

\section{Summary}
\label{sec:summary}
We have presented a Bayesian method for estimating the cross-spectrum
between two CMB temperature anisotropy maps.  The method is
advantageous for correlating maps because it does not require the maps
to have perfect overlap, identical beam shapes or pixelizations.
Using this formalism we found a high degree of correlation between the
maps from \maximaic, \wmapc, and \maximaii; in all cases the null
hypothesis is rejected with a probability higher than $1-10^{-8}$.
Additionally, we computed the power spectrum of the difference maps
for all combinations of the three data sets considered, and found that
in each case the spectra were consistent with the null spectrum.

The results show conclusively that the temperature fluctuations
detected by each of the \maximaic, \wmapc, and \maximaii experiments
are reproduced by these experiments, in overlapping regions of the
sky.  The close agreement of the fluctuations detected by these
experiments shows that current CMB experiments are now beginning to
provide us with high precision images of the true microwave sky.

\acknowledgments

All computations for this analysis were performed at the University of
Minnesota Supercomputing Institute in Minneapolis, Minnesota, and at
the National Research Scientific Computing Center in Berkeley,
California, which is supported by the Office of Science of the 
U.S. Department of Energy under contract no. DE-AC03-76SF00098.  
We acknowledge the use of the HEALPix software package.
MEA and RS acknowledge support from NASA grant no. S-92548-F.


\begin{thebibliography}{58}
\expandafter\ifx\csname natexlab\endcsname\relax\def\natexlab#1{#1}\fi

\bibitem[{{Abroe} et~al.(2002)}]{abroe/etal:2002}
Abroe, M.~E., et~al. 2002, \mnras, 334, 11

\bibitem[{{Bennett} et~al.(2003{\natexlab{a}})}]{bennett/etal:2003}
{Bennett}, C.~L., et~al. 2003{\natexlab{a}}, \apjs, 148, 1

\bibitem[{{Bennett} et~al.(2003{\natexlab{b}})}]{bennett/etal:2003b}
{Bennett}, C.~L., et~al. 2003{\natexlab{b}}, \apjs, 148, 175

\bibitem[{{Benoit} et~al.(2003)}]{benoit/etal:2003}
Benoit, A., et~al. 2002, A\&A, 399, L19

\bibitem[{{de Bernardis} et~al.(2000)}]{debernardis/etal:2000}
{de Bernardis}, P., et~al. 2002, Nature, 404, 955

\bibitem[{Bond, Jaffe, \& Knox(1998)}]{bjk:1998}
Bond, J.~R., Jaffe, A., \& Knox, L., 1998, \prd, 57, 2117


\bibitem[{{Ganga} et~al.(1993)}]{ganga/etal:1993}
Ganga, K., Page, L., Cheng, E., \& Meyer, S., 1993, \apjl, 509, L77

\bibitem[{Gorski, Hivon, \& Wandelt(1999)}]{healpix:1999}
Gorski, K.~M., Hivon, E., Wandelt B.~D., 1999, in Proceedings of the MPA/ESO 
Cosmology Conference "Evolution of Large-Scale Structure", 
eds. A.J. Banday, R.S. Sheth and L. Da Costa, PrintPartners Ipskamp, NL, pp. 37-42


\bibitem[{{Grainge} et~al.(2002)}]{grainge/etal:2002}
Grainge, K., et~al. 2002, submitted to \mnras

\bibitem[{{Halverson} et~al.(2002)}]{halverson/etal:2002}
Halverson, N.~W., et~al. 2002, \apj, 568, 38

\bibitem[{{Hanany} et~al.(2000)}]{hanany/etal:2000}
{Hanany}, S., et~al. 2000, \apj, 545, L5

\bibitem[{{Hinshaw} et~al.(2003)}]{hinshaw/etal:2003}
{Hinshaw}, G., et~al. 2003, \apj, submitted, astro-ph/0302217

\bibitem[{{Hinshaw} et~al.(2003)}]{hinshaw/etal:2003b}
{Hinshaw}, G., et~al. 2003, \apj, submitted, astro-ph/0302222

\bibitem[{{Hivon} et~al.(2002)}]{hivon/etal:2002}
{Hivon}, E., Gorski, K.~M., Netterfield, C.~B., Crill, B.~P.,
Prunet, S., \& Hansen, F., 2002, \apj, 567, 2

\bibitem[{{Jaffe} et~al.(2001)}]{jaffe/etal:2001}
{Jaffe}, A., et~al. 2001, \prl, 86, 3475

\bibitem[{{Jaffe} et~al.(2003)}]{jaffe/etal:2003}
{Jaffe}, A., et~al. 2003, submitted to \apj

\bibitem[{Kamionkowski et~al.(1996)Kamionkowski, Kosowsky, \& Stebbins}]{kam/etal:1996}
Kamionkowski, M., Kosowsky, A., \& Stebbins, A., 1996, \prd, 55, 7368

\bibitem[{{Knox} et~al.(1998){Knox}, {Bond}, {Jaffe}, 
{Segal}, \& {Charbonneau}}]{knox/etal:1998}
{Knox} L., {Bond} J.R., {Jaffe} A., {Segal} M., \&
{Charbonneau} D., 1998, \prd, 58


\bibitem[{Lee et~al.(2001)}]{lee/etal:2001}
Lee, A.~T. et~al. 2001, \apjl, 561, L1

\bibitem[{{Linweaver} et~al.(1995)}]{linweaver/etal:1995}
Linweaver, C.~H., et~al. 1995, \apj, 448, 482

\bibitem[{Mason et~al.(2003)}]{mason/etal:2003}
Mason, B.~S. et~al. 2003, \apj, 591, 540

\bibitem[{Hobson \& Maisinger(2002)}]{hm:2002}
Hobson, M. P., Maisinger, K., 2002, MNRAS, 334, 569

\bibitem[{Netterfield et~al.(2002)}]{netterfield/etal:2002}
Netterfield, B. et~al. 2001, \apj, 571, 604

\bibitem[{Padin et~al.(2001)}]{padin/etal:2001}
Padin, S. et~al. 2001, \apjl, 549, L1

\bibitem[{Press et~al.(1992)Press, Teukolsky, Vetterling, \&
  Flannery}]{press/etal:NRIC:2e}
Press, W.~H., Teukolsky, S.~A., Vetterling, W.~T., \& Flannery, B.~P. 1992,
  Numerical Recipes in C, 2nd edn. (Cambridge, UK: Cambridge University Press)

\bibitem[{Pryke et~al.(2002)}]{pryke/etal:2002}
Pryke, C., et~al. 2002, \apj, 568, 46



\bibitem[(Ruhl et~al.(2002)]{ruhl/etal:2002}
Ruhl, J.~E., et al., preprint, astro-ph/0212229 

\bibitem[{{Spergel} et~al.(2003{\natexlab{}})}]{spergel/etal:2003}
{Spergel}, D.~N., et~al. 2003{\natexlab{}}, \apj, submitted, astro-ph/0302209

\bibitem[{Stompor et~al.(2001)}]{stompor/etal:2001}
Stompor, R. et~al. 2001, \apjl, 561, L7

\bibitem[{Stompor et~al.(2002)}]{stompor/etal:2002}
Stompor, R. et~al. 2002, \prd, 65, 2003

\bibitem[{{Stompor} et~al.(2003)}]{stompor/etal:2003}
{Stompor}, R., et~al. 2003, C.R. Acad. Sci., Paris, t.0, Serie IV,
``The Cosmic Microwave Background:  Present
Status and Cosmological Perspectives''


\bibitem[{Tegmark(1999)}]{teg:1999}
Tegmark, M., 1999, \apjl, 519, 513

\bibitem[{Wu et~al.(2001)}]{proty/etal:2001}
Wu, J.H.P. et~al. 2001, \apjs, 132, 1


\bibitem[Xu, Tegmark, \& de Oliveira-Costa(2002)]{xu/etal:2002}
Xu, Y., Tegmark, M., \& de Olivier-Costa, A., 2002, \prd, 65, 083002 


\end{thebibliography}
\end{document}